\documentclass[aps,pra,twocolumn,preprintnumbers,amsmath,amssymb]{revtex4}
\usepackage{graphicx}
\usepackage{dcolumn}
\usepackage{bm}
\usepackage{paralist}

\newcommand{\ud}{\mathrm{d}}

\newcommand{\Reals}{\mathbb{R}}

\newcommand{\mb}[1]{\boldsymbol{#1}}
\newcommand{\br}{\mb{r}}
\newcommand{\bx}{\mb{x}}

\usepackage{slashed}

\newcommand{\braket}[2]{\langle{#1}|{#2}\rangle}

\newcommand{\brakket}[3]{\langle{#1}|{#2}|{#3}\rangle}

\newlength{\back}

\allowdisplaybreaks[2]

\begin{document}

\title{Asymptotic nodal planes in the electron density and the potential in the effective equation for the square root of the density} 
\author{Paola Gori-Giorgi}
\author{Evert Jan Baerends}

\affiliation{Theoretical Chemistry and Amsterdam Center for Multiscale Modeling, Vrije Universiteit, De Boelelaan 1083, 1081 HV Amsterdam, The Netherlands}

\date{\today}

\begin{abstract}
It is known that the asymptotic decay ($|\br|\to\infty$) of the electron density $n(\br)$ outside a molecule is informative about its first ionization potential $I_0$. It has recently become clear that the special circumstance that the Kohn-Sham (KS) highest-occupied molecular orbital (HOMO) has a nodal plane that extends to infinity may give rise to different cases for the asymptotic behavior of the exact density and of the exact KS potential [Mol. Phys. 114 (2016) 1086]. Here we investigate the consequences of such a HOMO nodal plane for the effective potential in the Schr\"odinger-like equation for the square root of the density, showing that for atoms and molecules it will usually diverge asymptotically on the plane, either exponentially or polynomially, depending on the coupling between Dyson orbitals. We also analyze the issue in the etxernal harmonic potential, reporting an example of an exact analytic density for a fully interacting system that exhibits a different asymptotic behavior on the nodal plane. 
\end{abstract}

\maketitle
\section{Introduction}
Both the (square root of the) density and the KS orbitals $\psi_k(\br)$ obey Schr\"odinger-type equations, 
\begin{eqnarray}
\label{eq:densityeqn}
\left(-\frac{1}{2}\nabla^2+v_{\rm ext}(\br)+v_{\rm eff}(\br)\right)\sqrt{n(\br)}  =   -I_0\sqrt{n(\br)}  \\
\label{eq:KSeqn}
\left(-\frac{1}{2}\nabla^2+v_{\rm ext}(\br)+v_{\rm Hxc}(\br)\right)\psi_k(\br)  = \epsilon_k\psi_k(\br),
\end{eqnarray}
where the sum of the external and the Hartree-exchange-correlation potentials constitutes the KS potential, $v_s=v_{\rm ext}+v_{\rm Hxc}$. The eigenvalue in Eq.~\eqref{eq:densityeqn} \cite{Hunter1975Symp9,LevyPerdewSahni1984} is the first ionization potential, $I_0=E_0^{N-1}-E_0^N$, and the occupied KS orbitals reproduce the density, $\sum_{k}^N|\psi_k(\br)|^2=n(\br)$. For the derivation of Eq.~\eqref{eq:densityeqn} it is essential to assume that the ground-state, interacting, $N$-electron wavefunction is real \cite{LevyPerdewSahni1984}. When this is not the case, an additional vector potential appears in the left-hand-side of Eq.~\eqref{eq:densityeqn}, as in the exact factorization approach put forward by Gross and coworkers \cite{AbeMaiGro-PRL-10,SchGro-PRL-17}. In what follows we only focus on the case in which the ground-state wavefunction is real, leaving the interesting investigation of the complex case to future work.

In molecules,  the external potential $v_{\rm ext}(\br)$ goes to zero at large distance like $-Z/r$, with $Z$ representing the total charge of all nuclei and $r$ the distance from the barycenter of nuclear charge. In this case, according to Eqs.~(\ref{eq:densityeqn})-(\ref{eq:KSeqn}), the asymptotic ($|\br|\to\infty$) decay of $\sqrt{n(\br)}$ and $\psi_k(\br)$ is (with $r=|\br|$) 
\begin{eqnarray}
\sqrt{n(\br)}	& \sim &  e^{-\sqrt{2(I_0+v_{\rm eff}(\infty))}\,r} \\
 \psi_k(\br) &	\sim & e^{-\sqrt{2(-\epsilon_k+v_{\rm Hxc}(\infty))}\,r}.
\end{eqnarray} 
Both the effective potential $v_{\rm eff}(\br)$ for $\sqrt{n(\br)}$ and the Hartree-exchange-correlation potential $v_{\rm Hxc}(\br)$ had been thought, until recently, to go to zero asymptotically. However, if a nodal plane extending to infinity is present in the Kohn-Sham highest-occupied molecular orbital (HOMO), a special behavior on this plane may result. Proposals for the asymptotic behavior of the exact $v_{\rm Hxc}(\br)$ have ranged from tending to a positive constant in the HOMO nodal plane (HNP) \cite{DellaSalaGoerling2002,DellaSalaGoerling2002b,KuemmelPerdew2003,KuemmelPerdew2003b} to a negative constant \cite{WuAyersYang2003}. The main question is whether the exact, interacting, density has a different decay on the HNP or whether this different decay is only a feature of a single-particle description, as this is what ultimately determines the exact $v_{\rm Hxc}(\br)$. A comprehensive investigation of this question has recently been performed for the exact density and the corresponding Hartree-exchange-correlation potential \cite{GoriGiorgiGalBaerends2016}, as well as for approximations like GGA's and metaGGA's \cite{Aschebrock2017}. For the exact density and KS potential, the analysis of the coupled equations for the Dyson orbitals $d_i(\br)$ did not allow for a final unique answer, but showed that two different scenarios are compatible with the structure of these equations. The simplest case might be a KS potential uniformly decaying like $-1/r$, even when the electron density has a different asymptotic decay in the HNP (namely as $\sim\exp[-2\sqrt{2I_1}\, r_p]$, with $I_1$ the second vertical ionization potential and $r_p$ a direction belonging to the HNP)  than elsewhere (where it is known to have asymptotics $\sim\exp[-2\sqrt{2I_0}\, r]$).  However, there are also cases where the density has exponential decay in the HNP according to $I_0$, like everywhere else. This will arise if the second Dyson orbital $d_1$  inherits this asymptotic behavior from the first Dyson orbital $d_0$ through angular coupling, and then necessarily the KS HOMO$-1$ will have to provide this same asymptotics in the HNP, since the KS HOMO does not contribute there.  In that case the KS potential will exhibit rather special behavior  \cite{GoriGiorgiGalBaerends2016} in order to induce the asymptotic decay in the HOMO$-1$ orbital different from the one according to its eigenvalue (which is typically close to the second ionization potential $I_1$). 

In the present paper we further investigate this issue focussing on the effective potential $v_{\rm eff}(\br)$ appearing in  Eq.~\eqref{eq:densityeqn}, which is related to the functional derivative of the von Weizs\"acker  kinetic energy functional \cite{Wei-ZP-35},
\begin{equation}
	T_{\rm W}[n]=\frac{1}{8}\int\frac{|\nabla n(\br)|^2}{n({\br})}d \br,
	\label{eq_TvW}
\end{equation}
via the relation
\begin{equation}
	v_{\rm eff}(\br)=\frac{\delta T_{\rm W}[n]}{\delta n(\br)}-v_{\rm ext}(\br)-I_0.
	\label{eq_relveffTvW}
\end{equation}
The functional $T_{\rm W}[n]$ is also often used in the construction of orbital-free kinetic energy functionals (see, e.g., Refs.~\onlinecite{Mar-PLA-86,LevOu--PRA-88,HolMar-PRA-91,LigCar-HMM-05}). The  corresponding energy density, $\tau_{\rm W}(\br)=\frac{|\nabla n(\br)|^2}{8 n({\br})}$, plays a crucial role for metaGGA functionals, where it is used to detect one-electron and iso-orbital regions (see, e.g., Refs.~\onlinecite{Becke1993,Bec-JCP-96,TaoPerStaScu-PRL-03,PerRuzTaoStaScuCso-JCP-05,ZhaSchTru-JCTC-06,PerRuzCsoConSun-PRL-09}).

As mentioned, the main open issue that ultimately determines the behavior of the exact KS and effective potential $v_{\rm eff}(\br)$ is whether a density coming from the wavefunction of a fully interacting system has or does not have a different asymptotic decay on the HNP. To further shed light on this open question, here we also report an interacting case that can be solved analytically (two spin-polarized electrons in the harmonic external potential), showing that its density displays different asymptotic decay on the HNP, and thus a different behavior of $v_{\rm eff}(\br)$, and discussing the implications for systems bound by the Coulomb potential.

The paper is organized as follows. In section \ref{Sec_asymptoticn} we will review some quantities needed in the discussion of the asymptotic behavior of the exact density. In particular, the definition of Dyson orbitals and the expansion of the exact density in a sum over the squares of the Dyson orbitals are relevant. Importantly, the asymptotic decay of the exact interacting density may be different in different directions: if there is a KS HOMO nodal plane, it may be inferred that also the first Dyson orbital (at eigenvalue $-I_0$) will have asymptotically the same nodal plane. In that case the decay $n(|\br_p|\to\infty)$ in that plane can be different (faster, according to the eigenvalue $I_1$ of the second Dyson orbital) than the decay outside the plane. But it might also happen that the second Dyson orbital inherits the slower decay on the plane from the first one, through the coupled equations \eqref{eq:Dysoneqn} for Dyson orbitals. In section \ref{sec:veffasymptotics} we give theoretical expressions for $v_{\rm eff}(\br)$, basically relating $v_{\rm eff}(\br)$ to wavefunction related quantities (such as the Dyson orbitals). We will recall that the KS potential can be expressed with the help of similar ingredients, with input from the KS independent particle wavefunction. The behavior of $v_{\rm eff}(\br)$  is then highlighted in sections \ref{sec_nonintCoul} and \ref{sec:harmonicpotential} using examples from both a Coulombic external potential $-Z/r$ and a harmonic external potential $\frac{1}{2}\omega^2 r^2$. The latter affords exact solutions, including electron correlation, for specific values of the $\omega$ \cite{Taut1993}, which will be used in section~\ref{sec_harminteract}. It is established that the potential $v_{\rm eff}(\br)$ of Eqs.~(\ref{eq:densityeqn}) and \eqref{eq_relveffTvW}, while normally going to zero asymptotically, exhibits a different asymptotic behavior in directions where the density decays differently: in the KS HNP it will usually diverge either exponentially or polynomially.  Conclusions are given in section \ref{Sec_Conclusions}. 

\section{Asymptotic behavior of the exact density}
\label{Sec_asymptoticn}
We briefly review a few aspects of the asymptotic behavior of the exact density based on the analysis of Ref.~\cite{GoriGiorgiGalBaerends2016}. It is known that the asymptotic behavior of the exact density of a molecule is related to its ionization energy \cite{KatrielDavidson1980}. The relation of the density to ion states can be made explicit with the so-called Dyson expansion of the wavefunction
\begin{eqnarray}
\Psi^N_0 &=& N^{-1/2}\sum_{i=0}^\infty d_i(\bx) \Psi^{N-1}_i(2 \cdots N), \nonumber  \\
d_i(\bx) & = & \sqrt{N} \int{ \Psi^{N-1}_i(2 \cdots N)^* \Psi^N_0(\bx,2 \cdots N)} \ud 2 \cdots \ud N,  \nonumber \\
n(\bx) & = & \sum_{i=0}^{\infty} |d_i(\bx)|^2,  \label{eq:Dysonorb}
\end{eqnarray}
where the $\Psi^{N-1}_i$ are the exact $(N-1)$-electron states and $\bx = \br,s$, with $s=\uparrow$ or $\downarrow$. Each state of the ion is associated with a one-particle wavefunction, its Dyson orbital.  The sum over $i$ goes over both the spin-$\uparrow$ and spin-$\downarrow$ Dyson orbitals. If e.g. $\bx=(\br,s=\uparrow)$ then only the spin-$\uparrow$ Dyson orbitals are nonzero at $\bx$ and contribute to $n(\bx)=n(\br,\uparrow)$ ($=\frac{1}{2}n(\br)$ in closed shell systems).
The Dyson orbitals constitute a nonorthogonal nonnormal, in general linearly dependent set. The Dyson orbitals are, however, not completely esoteric objects. In an independent particle system with a determinantal ground state wavefunction, such as the KS electrons, it follows from \eqref{eq:Dysonorb} that the Dyson orbitals are just the occupied orbitals (in this case there is only a finite number of nonzero Dyson orbitals). The expression of the density in terms of squares of Dyson orbitals is then equivalent to the KS expansion of the density in squares of KS orbitals.\\
 We define the conditional amplitude $\Phi(2 \cdots N;\bx)$ \cite{Hunter1975} and associated quantities, the conditional density $n^{cond}(\bx_2|\bx)$ and conditional potential $v^{cond}(\bx)$,
\begin{eqnarray}
\label{eq:condampl}
\Phi(2 \cdots N;\bx) & = & \frac{\Psi^N_0(\bx,2 \cdots N)} {\sqrt{n(\bx)/N} }, \nonumber \\
n^{cond}(\bx_2|\bx) & = & (N-1)\int |\Phi(2\cdots N|\bx)|^2 \ud 3\cdots \ud N, \nonumber  \\
v^{cond}(\bx) & =& \int{ \frac{n^{cond}(\bx_2|\bx)}{|\br-\br_2|} } \ud \bx_2.
\end{eqnarray}
$\Phi(2 \cdots N;\bx)$ is a normalized $(N-1)$-electron wavefunction depending parametrically on the position $\bx$. Its square describes the probability distribution of electrons at positions $2 \cdots N$ when one electron is known to be at $\bx$. Its associated one-electron density $n^{cond}(\bx_2|\bx)$ is the density of the other electrons at position $\bx_2$ when one electron is at $\bx$, which is the normal one-electron density $n(\bx_2)$ plus the full exchange-correlation hole surrounding position $\bx$. 
Projecting the Schr\"odinger equation $\hat{H}^N \Psi^N_0=E_0^N \Psi_0^N$ against $\Psi_i^{N-1}(2 \cdots N)$ and using the expansion of Eq.~\eqref{eq:Dysonorb} one obtains the usual equations for the Dyson orbitals,
\begin{equation}
\label{eq:Dysoneqn}
\left(-\frac{1}{2}\nabla^2+v_{\rm ext}(\br)\right)d_i(\bx) +\sum_{k=0}^\infty X_{ik}(\bx)d_k(\bx) =  -I_id_i(\bx).
\end{equation}
Katriel and Davidson (KD) \cite{KatrielDavidson1980} pointed out that, due to the coupling integrals
\begin{equation}\label{eq:coupling}
  X_{ik}(\bx)\equiv\brakket{\Psi_i^{N-1}}{\sum_{j > 1}^N\frac{1}{|\br_j-\br|} }{\Psi_k^{N-1}}_{2..N},
\end{equation} 
the exponential decay of the coupled Dyson orbitals will be the same, cf.\ Ref.~\cite{HandyMarronSilverstone1969} for the case of Hartree-Fock orbitals. The first Dyson orbital $d_0$ will have exponential decay $\sim e^{-\sqrt{2I_0}\,r}$ multiplied by a factor $r^{\beta}$ with $\beta=(Z-N+1)/\sqrt{2 I_0}-1$, due to the $-Z/r$ decay of $v_{\rm ext}$ and the $(N-1)/r$ decay of the coupling term. KD find that higher Dyson orbitals which have nonzero $X_{i0}$ with the first Dyson orbital will have decay $r^{\beta-L^*}e^{-\sqrt{2I_0}\,r}$ with $L^* \ge 2$. Dyson orbitals that are not connected to  $d_0$ will have different exponential  decay, governed by the eigenvalue of the first orbital in such a connected set (which is disjunct from other sets).
Considering the expansion of the density in Dyson orbitals in Eq.~\eqref{eq:Dysonorb}, KD have concluded that, if the density decays for $|\br| \to \infty$ as the most slowly decaying term $|d_0(\bx)|^2$, its exponential decay (we do not write here the polynomial prefactor) would be $\sim e^{-2\sqrt{2I_0}\,r}$. Levy, Perdew and Sahni (LPS) \cite{LevyPerdewSahni1984} proved this exponential decay in a different way, thereby showing that the leading term is not overruled by the infinite sum of the faster decaying terms in Eq.~\eqref{eq:Dysonorb}. The result 
\begin{equation}\label{eq:nasymptotic}
n(|\br|\to\infty) \sim |d_0(\br)|^2 \sim e^{-2\sqrt{2I_0}\,r} 
\end{equation}
is generally accepted.

It has been realized \cite{DellaSalaGoerling2002,KuemmelPerdew2003,WuAyersYang2003,Holas2008} that there may be special cases where the asymptotic behavior of the density is different in some directions than the one of Eq.~\eqref{eq:nasymptotic}. This may happen, for instance, when there is a symmetry plane in the system (as in many $\pi$ systems, like ethylene and benzene), but also in more general situations.  In the case of a symmetry plane, the exact interacting states of the molecule are either symmetric or antisymmetric with respect to the plane. For example, if the KS HOMO has a nodal plane, the ground state KS wavefunction (and very likely also the exact ground state wavefunction) corresponding to a closed shell configuration is totally symmetric with respect to that plane, while the first ion state will be antisymmetric (the KS first ion state surely will be so, and  we will consider the usual case that the same holds for the exact ion state.) For points $\br_p$ in the HNP the conditional amplitude $\Phi(2 \cdots N|\bx_p)$ will be symmetric with respect to the plane. Therefore, the matrix element  $\braket{\Psi_0^{N-1}}{\Phi(2 \cdots N|\bx_p)}_{2..N}$ will vanish, so that the first Dyson orbital is zero in the plane:
\begin{align}
\label{d0} 
&d_0(\bx_p)=\sqrt{n(\bx_p)} \braket{\Psi_0^{N-1}}{\Phi(2 \cdots N|\bx_p)}_{2..N}=0. 
\end{align}
In fact, $d_0$ is antisymmetric with respect to the plane.  When we consider the asymptotic behavior of higher Dyson orbitals, it is clear that with $d_0(\bx_p)=0$, the coupling to $d_0$ in 
Eq.~\eqref{eq:Dysoneqn} for points in the HNP at first sight seems to be zero for any higher Dyson orbital $d_{i>0}$ (but see below). The decay in the HNP of the second Dyson orbital (and thus of the density) is then not governed by $d_0$ but by $d_1$ with asymptotic behavior according to exp$[-\sqrt{2I_1}\, r_p]$. This is what has been called Case 1 in Ref.~\cite{GoriGiorgiGalBaerends2016}. It is exemplified by the minimal model for a density employed by Aschebrock et al.\ \cite{Aschebrock2017}, with a $p_z$ type orbital with (outside the HNP $z=0$) slow decay $\sim\exp[-\alpha_p r]$ and a lower lying $s$-type orbital with faster decay $\sim\exp[-\alpha_s r]$, $\alpha_s > \alpha_p$ (cf. our $\sqrt{2I_0}\,r$ and $\sqrt{2I_1}\,r$ for the exponents of HOMO and HOMO$-1$ respectively). Ref.~\cite{Aschebrock2017} gives a comprehensive discussion of the shape of exchange potentials obtained as functional derivatives of GGA exchange energy approximations (Armiento-K\"ummel \cite{ArmientoKuemmel2013} and B88 \cite{Becke1988}), as well as potential functionals like Becke-Johnson \cite{BeckeJohnson2006} and LB94 \cite{LeeuwenBaerends1994}. In that investigation the minimal model density is fed into the density  functionals for the various potentials. Often an exponential diverging behavior is obtained of the form $\exp[k(\alpha_s-\alpha_p)r]$ ($k=1$ or $1/2$). Remarkably, the same exponential divergence has been observed \cite{GoriGiorgiGalBaerends2016} for the effective potential for the square root of the density for an exact density like the minimal model. Such a density with different decay in a particular plane than elsewhere has been called Case 1 \cite{GoriGiorgiGalBaerends2016} (fast decay according to $I_1$ in the plane, slower decay according to $I_0$ everywhere else). However, it is an important issue whether a true density of Coulombically interacting electrons can have such different exponential decay in different directions. The present authors have argued that an exact density will typically not exhibit this different exponential decay in different directions (although the polynomial prefactor may differ). This has been called Case 2 in \cite{GoriGiorgiGalBaerends2016}. Decay of the HOMO$-1$ according to $I_0$ in the HNP leads to rather intricate consequences for the KS potential, which requires very special features to generate a decay of HOMO$-1$ in HNP according to $I_0$ and not according to its eigenvalue (which is equal to (or close to) $I_1$). The situation for $v_{\rm eff}(\br)$ is, however, simpler than for the KS potential, since it can be related directly to wavefunction quantities, as discussed in section \ref{sec:veffasymptotics}. \\

\section{Asymptotic behavior of the effective potential for $\sqrt{n}$} 
\label{sec:veffasymptotics}

The potential $v_{\rm eff}(\br)$ of Eq.~\eqref{eq:densityeqn} can be written in the form \cite{LevyPerdewSahni1984,BuijseBaerendsSnijders1989}
\begin{eqnarray}\label{eq:veff}
v_{\rm eff}(\bx) & = &  v^{cond}(\bx)+\frac{1}{2} \braket{\nabla_x\Phi(2 \cdots N|\bx) }{ \nabla_x\Phi(2 \cdots N|\bx)}   \nonumber \\
& + & \brakket{\Phi(2 \cdots N|\bx)}{\hat{H}^{N-1}-E^{N-1}_0}{\Phi(2 \cdots N|\bx)} \nonumber \\
		&\equiv & v^{cond}(\bx) +v^{kin}(\bx) \label{eq:veffeqn} + v^{N-1}(\bx).
\end{eqnarray}
For future reference we note that for the exact KS potential an analogous expression holds,
\begin{equation}
\label{eq:vs}
v_s=v_{\rm ext}+v^{cond}+(v^{kin}-v_s^{kin})+(v^{N-1}-v_s^{N-1}).
\end{equation}
The potentials $v_s^{kin}$ and $v_s^{N-1}$ depend on the KS independent particle wavefunction $\Psi_s^N$ and notably its associated conditional amplitude $\Phi_s$ in exactly the same way as $v^{kin}$ and $v^{N-1}$ depend on the exact wavefunction and conditional amplitude.
In Eq.~\eqref{eq:veff} $v_{\rm eff}(\br)$ is expressed in terms of only wavefunction quantities. LPS \cite{LevyPerdewSahni1984} stressed that each term in Eq.~\eqref{eq:veffeqn} is everywhere nonnegative and should tend to zero asymptotically. In fact, $v^{cond}(\bx)$ [see Eq.~\eqref{eq:condampl}], being the repulsive Coulomb potential of a localized charge distribution of $(N-1)$ electrons, decays like $(N-1)/r$.  
The third term of $v_{\rm eff}$, $v^{N-1}$, is positive since in general $\Phi$ will not be the ground state wavefunction of the ion, so its expectation value will be larger than $E_0^{N-1}$. When $|\br| \to \infty$ it has been inferred \cite{KatrielDavidson1980}  that the conditional amplitude collapses to the ion ground state $\Psi_0^{N-1}$ (when $s=\uparrow$ then $\Phi$ will collapse to the $M_S=-1/2$ state of the doublet ion), so that $v^{N-1}(|\br|\to \infty)\to 0$. The second term, $v^{kin}$, is manifestly positive and is expected to go to zero asymptotically since the derivative of $\Phi$ with respect to $\br$ when the reference electron is very far becomes zero  ($\Phi$ remains constant -- the ion ground state -- under small change of $\br$ at $\infty$). These expectations are not borne out if there is a HNP, see below. \\
\subsection{Case 1: The density decay on the HNP is governed by the HOMO$-1$}\label{subsec:Case1}
For points $\br_p$ in the HNP $d_0(\bx_p) = 0$ because of spatial symmetry.
By expanding the conditional amplitude $\Phi(2 \cdots N|\bx)$  in terms of the exact $N-1$ states,
\begin{equation}
 \Phi(2 \cdots N|\bx)=\sum_{i=0}^\infty \frac{d_i(\bx)}{\sqrt{n(\bx)}} \Psi^{N-1}_i(2 \cdots N),  
\end{equation}
we see that, since on the HNP $d_0=0$ and $|d_1(\bx_p)| (r_p \to \infty) \sim \sqrt{n(\bx_p)}$, while all higher $d_i$ decay a factor $r^{-L^*}$ faster \cite{KatrielDavidson1980}, with $L^* \ge 2$,  the conditional amplitude tends asymptotically on the plane to the first-excited ion state, $\Phi\to \Psi_1^{N-1}$ (note that for any position $\bx$ $\Phi$ is normalized).  This implies that
\begin{equation}
\label{eq:vN-1}
	v^{N-1}(|\br_p|\to\infty)=E_1^{N-1}-E_0^{N-1}=I_1-I_0.
\end{equation}
This is a {\em positive} constant appearing in the asymptotics of $v_{\rm eff}$ only on the HNP. The exponential decay of $\sqrt{n}$ is governed according to Eq.\ \eqref{eq:densityeqn} by $\exp(-\sqrt{2(I_0+v_{\rm eff}(\infty))}r)$.  The positive asymptotic constant in $v_{\rm eff}(\br \to \infty)$ looks perfectly in order: this value for $v_{\rm eff}(\infty)$ gives precisely the asympotic decay $\exp(-\sqrt{2I_1}r_p)$ we have assumed for $\sqrt{n}$ on the HNP in Case 1. 

Also the second term in Eq.~\eqref{eq:veffeqn}, $v^{kin}$, can be nonzero at infinity: when crossing the HNP, the asymptotic conditional amplitude changes from $\Psi_0^{N-1}$, to which it collapses for asymptotic points in general directions,  to $\Psi_1^{N-1}$, to which it collapses for asymptotic points in HNP, see above.  So the $\br$-derivative of $\Phi$ perpendicular to the plane can be nonzero on the HNP also when $|\br|\to\infty$. Its actual value depends on how $d_0(\br\to\br_p)$ goes to zero when approaching the nodal plane. We have noted that for the determinantal wavefunction of a noninteracting system the Dyson orbitals are precisely the occupied independent particle orbitals. In the interacting system the first Dyson orbitals for primary ion states (those corresponding to a simple orbital ionization) still are very similar to the Kohn-Sham orbitals: overlaps are typically $> 0.999$ \cite{GritsenkoBraidaBaerends2003}. This agrees with our finding in this paper that when the KS HOMO is antisymmetric with respect to a plane, the corresponding Dyson orbital also is antisymmetric with respect to that plane. Let us then take as example that asymptotically, in spherical coordinates,  $d_0\sim f(\cos\theta)R(r)e^{-\sqrt{2I_0}\,r}$, with $f(0)=0$, and $f'(0)\ne0$, as would be the case for a $\pi$ orbital,  which has $fR=r\cos\theta=z$.  By writing $v^{kin}$ in the 
form~\cite{BuijseBaerendsSnijders1989,BaerendsGritsenkoJPCA1997,ChongGritsenkoBaerends2002}
\begin{equation}
\label{eq:vkin}
	v^{kin}(\br)=\frac{1}{2}\sum_{i=0}^\infty\frac{|\nabla d_i(\bx)|^2}{n(\br)}-\frac{|\nabla n(\br)|^2}{8 n(\br)^2},
\end{equation}
 and using $d_1\sim e^{-\sqrt{2I_1}\,r}$, it is found after some manipulation that 
\begin{equation}
v^{kin}(r_p\to\infty)\to \frac{1}{2}f'(0)^2\frac{R^2}{r^2}e^{2(\sqrt{2I_1}-\sqrt{2I_0})\,r},
\label{eq:vkinasym}
\end{equation}
showing that $v^{kin}$ can go asymptotically to infinity on the HNP. 
A simple illustration of this fact is given in the next Sec.~\ref{sec_nonintCoul} for non-interacting electrons (a Case 1 density).  

The asymptotically diverging behavior of Eq.~\eqref{eq:vkinasym} is perfectly compatible with an analytical, well-behaved density. It induces in the density the special behavior in the  HNP of Case 1 which is certainly realizable by noninteracting electrons in one-electron states (orbitals): fast decay according to $I_1$ in HNP coming from HOMO$-1$, slow decay everywhere else according to $I_0$ from HOMO. The key point is that when we project Eq.~\eqref{eq:densityeqn} on the plane, we have to take into account also $\nabla^2\sqrt{n}$ in the direction perpendicular to the plane. Usually, the $\theta$ and $\phi$ derivatives in $\nabla^2$ are zero when $r\to\infty$, but when there is a HNP this is not the case. Indeed in our example, using spherical coordinates, the $-\frac{1}{r^2\sin\theta}\frac{\partial}{\partial \theta}\left(\sin\theta\frac{\partial}{\partial \theta}\right)$ operation on $\sqrt{n}$ exactly cancels the diverging behavior coming from $v_{kin}$. This leaves for $r_p \to \infty$ just the radial part of the one-electron Schr\"odinger equation. With the remaining potential $-1/r$ from $v_{ext}+v^{cond}$, and the $I_1-I_0$ constant of $v^{N-1}$ combined with the eigenvalue $-I_0$, $\sqrt{n}$ acquires the asymptotic decay in the HNP according to $I_1$.\\  
For an interacting electron system the density is  described by the leading terms in the Dyson expansion, $n(\bx)=|d_0(\bx)|^2+|d_1(\bx)|^2+\dots$. Only if there is no coupling of $d_{i>0}$ to $d_0$ in Eq.~\eqref{eq:Dysoneqn} will $d_1$ (and the orbitals in the same set) have asymptotics according to $I_1$ and will this picture for noninteracting electrons also prevail for the interacting electron system. We discuss in the next subsection the Case 2 where such coupling does occur.  \\
Considering the kinetic correlation potential $v_c^{kin}=v^{kin}-v_s^{kin}$ in the KS potential, we have argued in Ref.\ \cite{GoriGiorgiGalBaerends2016} that if the exact density is like the noninteracting (KS) density with a HOMO nodal plane, the behavior of the Dyson orbitals $d_0$ and $d_1$ close to HNP should be identical to that of HOMO and HOMO$-1$, and in $v_c^{kin}$ the divergence of $v^{kin}$ is canceled by an equal divergence of $v_s^{kin}$. Then in Case 1 the KS potential will have asymptotically the simple uniform $-1/r$ behavior, compatible with solutions of the KS equations with a HOMO with a nodal plane and a HOMO$-1$ with uniformly faster decay (the density of the minimal model of Aschebrock et al.\ \cite{Aschebrock2017} is compatible with such a regular KS potential).  This is then a consistent picture. However, we have also indicated that the situation where coupling of $d_1$ to $d_0$ in Eq.~\eqref{eq:Dysoneqn} generates slow decay in $d_1$ will be prevalent in interacting electron systems, see discussion of Case 2 in next section.

As recalled in Eq.~\eqref{eq_relveffTvW}, the potential $v_{\rm eff}(\br)$ essentially gives the functional derivative of the von Weizs\"acker  kinetic energy functional, which, thus, also exhibits the same diverging behavior on the nodal plane in Case 1. Notice that all the spherical harmonics $Y_{\ell m}(\theta,\phi)$ approach their nodal planes linearly with $\cos \theta$, so that the divergence predicted by Eq.~\eqref{eq:vkinasym} is expected to occur in the general case. This might have consequences for calculations using orbital-free kinetic energy functionals and metaGGA functionals, probably in a way qualitatively similar to the one reported by Aschebrock et al.~\cite{Aschebrock2017}.\\

\subsection{Case 2: The density decay on HNP is exponentially the same as everywhere, although polynomially faster}\label{subsec:Case2}
In Ref.~\cite{GoriGiorgiGalBaerends2016} it has been shown that in Coulombically interacting systems coupling of some of the $d_{i>0}$ to $d_0$ in Eq.~\eqref{eq:Dysoneqn} will usually occur (Case 2), and will lead to $d_1$ (and other orbitals) acquiring on the nodal plane the same slow exponential asymptotic decay (dictated by $I_0$) as $d_0$ in general directions. The decay of $d_1$ will then still be polynomially faster on the plane (by $1/r^4$) (and the correlated density a factor $1/r^8$ faster).  We have investigated what the asymptotic behavior of $v^{kin}$ will be for such a Case 2 density. Let us consider the essential terms in $d_1$ responsable for the slow $e^{-\sqrt{2I_0}\,r}$ behavior \cite{GoriGiorgiGalBaerends2016}, notably also the term  $Ce^{-\sqrt{2I_0}\,r}/r^3$ yielding this slow decay of $d_1$ on the plane
\begin{align}
d_0 &\sim q_1 r\cos{\theta} e^{-\sqrt{2I_0}\,r} \notag \\
d_1 &\sim \frac{f(\cos\theta)}{r} e^{-\sqrt{2I_0}\,r} + \frac{C_1}{r^3} e^{-\sqrt{2I_0}\,r} + D_1 r^n e^{-\sqrt{2I_1}\,r} \notag \\
& \text{ with }f(0)=0 \notag. \\
\rho(r,\theta) &= |d_0(r,\theta)|^2 + |d_1(r,\theta)|^2,
\end{align}
where the constant $C_1$ is non-zero if $f''(0)\neq 0$, which usually will occur, since, as discussed in Ref.~\cite{GoriGiorgiGalBaerends2016}, in the vast majority of cases we will have $f(x)=x^2$.
After some manipulation one obtains for the asymptotics of $v^{kin}$
\begin{align}\label{eq:vkinasym2}
v^{kin}(r&\to \infty,\theta =\frac{\pi}{2}) = \frac{q_1^2 r^6}{2 (C_1+D_1e^{(\sqrt{2I_0}-\sqrt{2I_1})r} r^{3+n})^2} \notag \\
             & \to \frac{q_1^2 r^6}{2 C_1^2}
\end{align} 
So $v^{kin}$ will still be diverging on the nodal plane of the first Dyson orbital $d_0$, but the divergence is no longer exponential, but becomes polynomial (like $r^6$). Note that if $C_1 \to 0$, i.e.\ when $d_1$ does not have the slow decay on the plane, then we are back in Case 1 above, where $v^{kin}$ diverges more rapidly on the plane, in fact exponentially as in Eq.~\eqref{eq:vkinasym}. One may again verify that the $r^6$ divergent behavior does not pose any problem in Eq.~\eqref{eq:densityeqn} for $\sqrt{n}$ since it is canceled by an opposite divergent term coming from the Laplacian of $\sqrt{n}$. 

A more detailed analysis including all the Dyson orbitals $d_{i>0}$ that inherit the slow decay on the HNP from $d_0$ through the same kind of angular coupling does not change qualitatively the conclusion of Eq.~\eqref{eq:vkinasym2}, with the {\it caveat} that one should always be careful with asymptotic expansions expressed as infinite sums. Equation \eqref{eq:vkinasym2}, in fact, becomes
\begin{equation}
	v^{kin}(r \to \infty,\theta =\frac{\pi}{2})\to \frac{q_1^2 r^6}{2 \sum_{i\in \mathcal{G}}^\infty C_i^2},
\end{equation}
where $i\in \mathcal{G}$ denotes the set of all the Dyson orbitals having the same coupling with $d_0$ as $d_1$ (this set includes all the Dyson orbitals for which the matrix element $k_i$ appearing in Eq.~(16) of Ref.~\cite{GoriGiorgiGalBaerends2016} is nonzero, the constants $C_i$ being determined by the matrix element $k_i$ divided by $(I_i-I_0)^2$. In highly symmetrical systems, like ethylene and benzene, many $k_i$ will be zero by symmetry \cite{GoriGiorgiGalBaerends2016}.

Also, one should keep in mind that now the conditional probablity on the HNP does not collapse asymptotically anymore to the first excited state of the ion, but to a superposition of all the ion states  with Dyson orbitals that are non-zero on the plane (belonging to the set $\mathcal{G}$ having $C_i \ne 0$),
\begin{equation}
	\Phi(2\dots N, |\br_p|\to\infty)\to \frac{\sum_{i\in \mathcal{G}}^\infty C_i^2 \Psi_i^{N-1}(2\cdots N)}{\sqrt{\sum_{i\in \mathcal{G}}^\infty C_i^2}}.
\end{equation}
As a consequence, Eq.~\eqref{eq:vN-1} does not hold anymore and we have, instead,
\begin{equation}
	v^{N-1}(|\br_p|\to\infty)\to \frac{\sum_{i\in \mathcal{G}}^\infty \frac{k_i^2}{(I_i-I_0)^3}}{\sum_{i\in \mathcal{G}}^\infty \frac{k_i^2}{(I_i-I_0)^4}},
\end{equation}
which reduces to Eq.~\eqref{eq:vN-1} when only $i=1$ is considered.

\subsection{External harmonic potential}\label{subsec:Harmonium}

We also analyze how these conclusions may become different for a different external potential. We take the case of an harmonic external potential, $v_{ext}=\frac{1}{2}\omega^2r^2$, which has the interesting property that it affords analytic solutions for two Coulombically interacting electrons for specific values of $\omega$ \cite{Taut1993}. Moreover, just as the Coulombic external potential, it has many applications in physics (quantum dots, cold atoms, plasmas, etc.). Equation \eqref{eq:densityeqn} is generally valid for any binding external potential, including the harmonic confinement. From the decomposition of  $v_{\rm eff}(\br)$ of Eq.~\eqref{eq:veffeqn} we clearly see that also in this case $v_{\rm eff}(\br)$ is expected to go asymptotically to zero. In other words, it is only the Coulombic nature of the electron-electron repulsion, determining the conditional amplitude and related quantities, that matters for the asymptotic behavior of $v_{\rm eff}(\br)$. It is again a special behavior of the density in certain directions, such as in a HNP, that may induce special behavior of $v_{\rm eff}(\br)$ in these directions.

Obviously, the way the information on $I_0$  is embodied into the asymptotics of the density  is different with the harmonic external potential than with the Coulombic external potential of the molecular case.  
 In $d$ dimensions, with $v_{\rm eff}(|\br|\to\infty)\to 0$, Eq.\ \eqref{eq:densityeqn} for the square root of the density of $N$ electrons confined in an harmonic trap reads, when $r\to\infty$,
\begin{equation}
	\left(-\frac{1}{2}\nabla^2+\frac{1}{2}\omega^2 r^2\right)\sqrt{n(\br)}=\left(E_0^N-E_0^{N-1}\right)\sqrt{n(\br)}.
\label{eq_densharmasy}
\end{equation}
As well known, the solution has the form $e^{-\frac{\omega}{2}r^2}u(r)$, where for large $r$, $u(r)\sim r^q$, with $q\in \Reals^+$. While the gaussian decay only depends on $\omega$, it is now the polynomial prefactor ($q$) that carries the information on $E_0^N-E_0^{N-1}$,
\begin{equation}
 E_0^N-E_0^{N-1}=\omega\left(\frac{d}{2}+q\right) \qquad q\in\Reals^+, 
\end{equation}
with
\begin{equation}
	\sqrt{n(|\br|\to\infty)}\sim e^{-\frac{\omega}{2}r^2}r^q,
\end{equation}
and hence  in three dimensions ($d=3$)
\begin{eqnarray}
&n(|\br|\to\infty)\sim r^{2 q} e^{-\omega r^2} \qquad   q=\left(-\frac{I_0}{\omega}-\frac{3}{2}\right)\in \Reals^+ \notag  \\
 &-I_0 = E_0^N-E_0^{N-1}
\label{eq:asymptHarm}
\end{eqnarray}
Notice that, due to the unbounded nature of the potential, the ionization energy cannot be defined as removing the particle to infinity with zero kinetic energy. Removing a particle (ionization) is equivalent to putting the particle with zero kinetic energy at the zero of the harmonic potential well (without interaction with the other particles). The energy of a $(N-1)$-particle state in the harmonic potential $minus$ the energy of a $N$ particle state is then negative with $-I_0\ge \frac{3}{2}\omega$, so that $q$ is positive. The negative ionization energy does not change the derivation of Eq.~\eqref{eq:densityeqn}. 
The exact solutions that are possible in this case afford an analytical study of the asymptotic behavior of $v_{\rm eff}(\br)$ for both a noninteracting and an interacting correlated electron system in the presence of a KS HOMO nodal plane, as reported in section~\ref{sec:harmonicpotential}.

We should note, however, that because the ionization information is now only in the polynomial prefactor, Cases 1 and 2 discussed for the external Coulomb potential in sections~\ref{subsec:Case1} and \ref{subsec:Case2}, respectively, can get mixed in the harmonic external potential. The reason is that, as explained in section~\ref{subsec:Case2}, when the angular coupling between the Dyson orbitals makes $d_1$ inherit the slower behavior of $d_0$ on the plane, such behavior is damped by a factor $1/r^4$ (which becomes $1/r^8$ in the density). This polynomial damping does not prevent $d_1$ from getting a slower decay on the plane if the difference in the two asymptotic behaviors is exponential, as it happens for the Coulomb external potential case, but can have an important effect when the difference is only polynomial.


\section{Non-interacting electrons in the Coulomb external potential}
\label{sec_nonintCoul}

In order to illustrate the diverging behavior predicted by Eq.~\eqref{eq:vkinasym}, we consider $N=3$ non-interacting electrons in the Coulomb external potential $v_{\rm ext}(\br)=-Z/r$ in the configuration $1s^22p_z^1$, which is one of the possible degenerate ground-states.  The corresponding wavefunction is antisymmetric with respect to the nodal plane $z=0$. We construct the corresponding density and we calculate $v_{\rm eff}(\br)$ by inversion,
\begin{equation}
v_{\rm eff}(\br)=\frac{\nabla^2\sqrt{n(\br)}}{2\sqrt{n(\br)}}-v_{\rm ext}(\br)-I_0.
\label{eq:veffbyinversion}
\end{equation}
\begin{figure}
   \includegraphics[width=8cm]{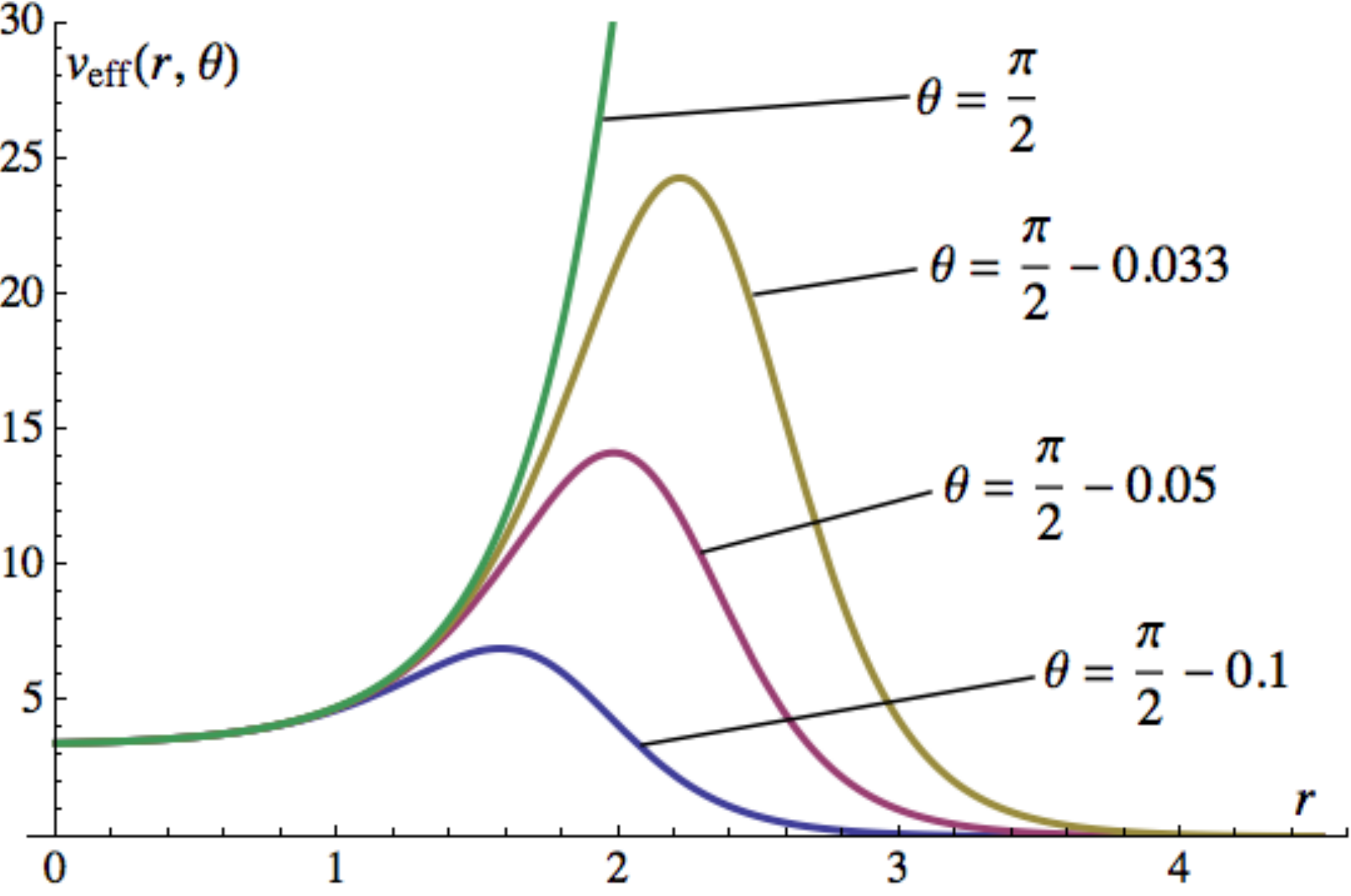}
\includegraphics[width=8cm]{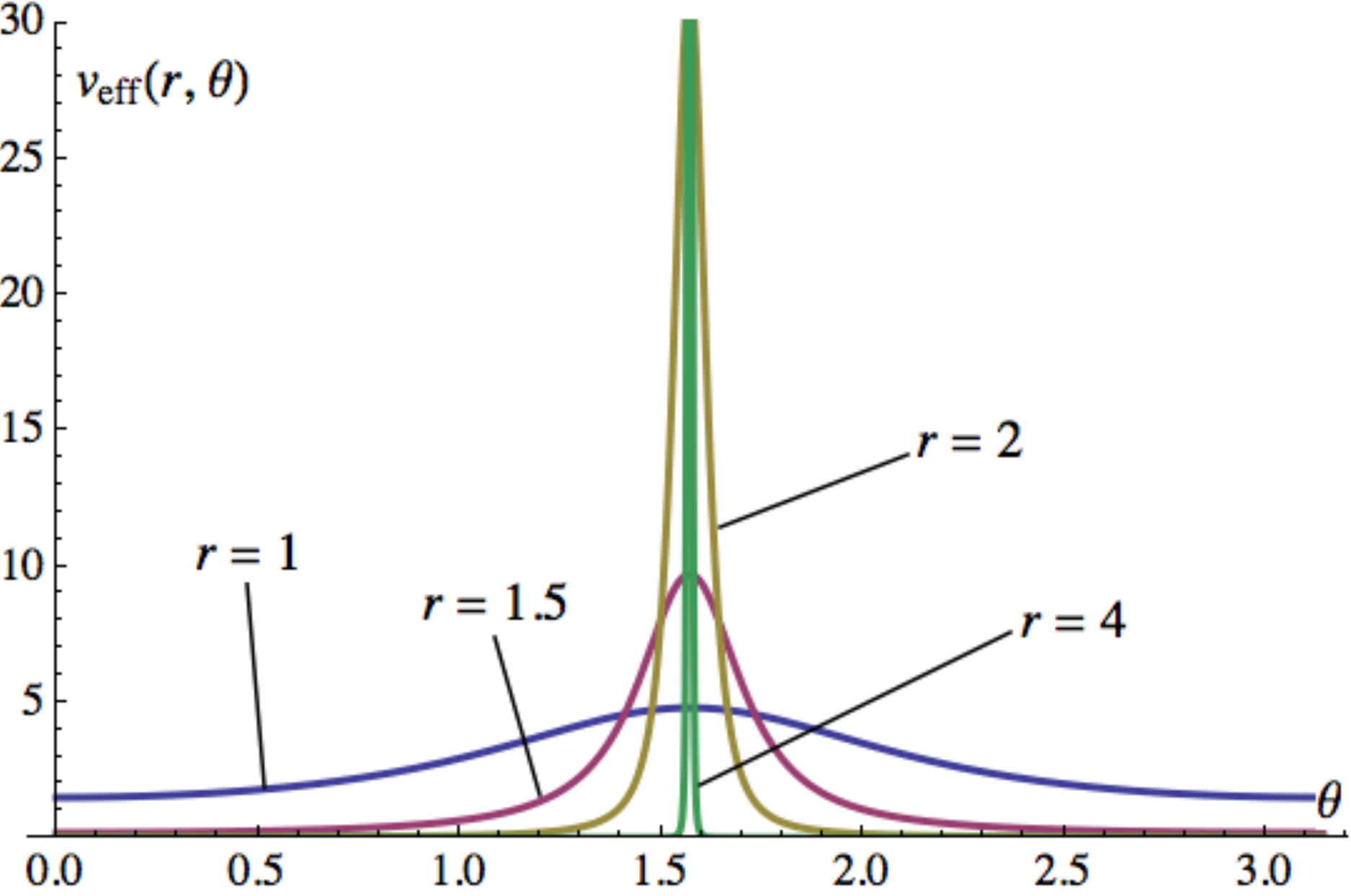}
   \caption{The effective potential $v_{\rm eff}(\br)$ for $\sqrt{n(\br)}$ in the case of $N=3$ non-interacting electrons in the external potential $v_{\rm ext}(\br)=-3/r$ in the configuration $1s^22p_z^1$. Top panel: $v_{\rm eff}(\br)$ as a function of $r=|\br|$ for different values of $\theta=\arccos(\frac{z}{r})$. Bottom panel: $v_{\rm eff}(\br)$ as a function of $\theta$ for different values of $r$. }
\label{fig:veff3ele}
\end{figure}
The result is shown in Fig.~\ref{fig:veff3ele} for $Z=3$  as a function of $r$ for different values of $\theta=\arccos(\frac{z}{r})$ (top panel) and as a function of $\theta$ for different values of $r$ (bottom panel). We see that $v_{\rm eff}(\br)$ is a smooth function going to zero asymptotically everywhere except on the plane $z=0$, where it diverges exponentially according to Eq.~\eqref{eq:vkinasym}.

The diverging behavior we illustrate here for $v_{\rm eff}(\br)$ is precisely the same as the one obtained for various approximations (like GGAs) to the KS potential by Aschebrock et al.\ \cite{Aschebrock2017}, if a density like the present one is inserted in the corresponding exchange potential expressions. However, such a density (called minimal model in \cite{Aschebrock2017}) is generated here by a purely Coulombic potential  in the Schr\"odinger equation \eqref{eq:KSeqn}. It is not clear if the similarities between the present exact   $v_{\rm eff}(\br)$ potential for this type of $\sqrt{n}$ (with faster decay on the HNP) and the approximate (GGA) KS potentials obtained with this density are more than accidental.

\section{Electrons in a harmonic external potential}
\label{sec:harmonicpotential}

We illustrate here the behavior of $\sqrt{n}$ and $v_{eff}(\br)$ in the harmonic external potential in the two cases of non-interacting and interacting electrons, where we use one of the analytic solutions of Taut \cite{Taut1993} for the spin-polarized (triplet) case.

\subsection{Non-interacting electrons}
\label{sec_harmnoninter}
\begin{figure}
   \includegraphics[width=8.cm]{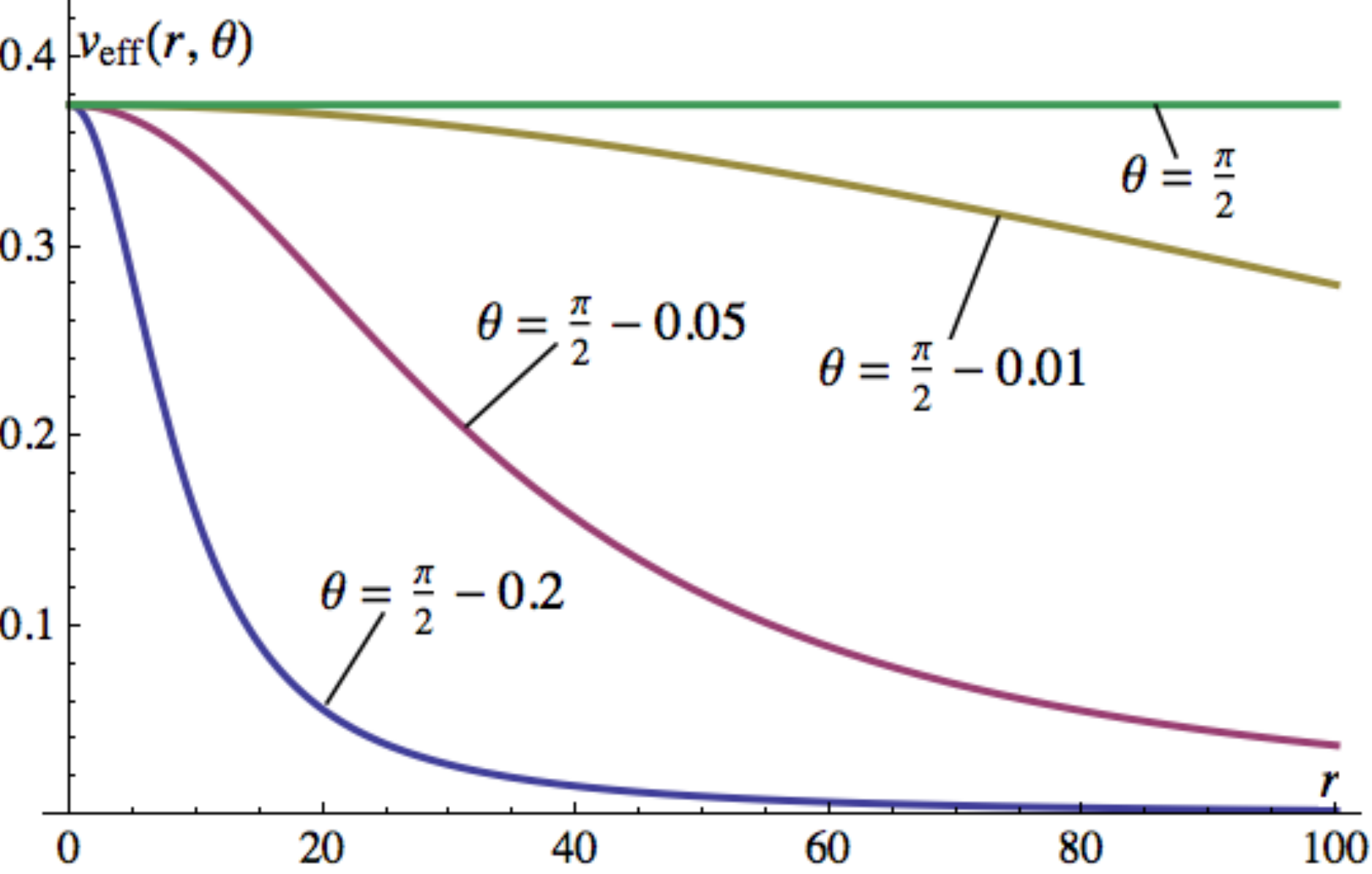}
\includegraphics[width=8.cm]{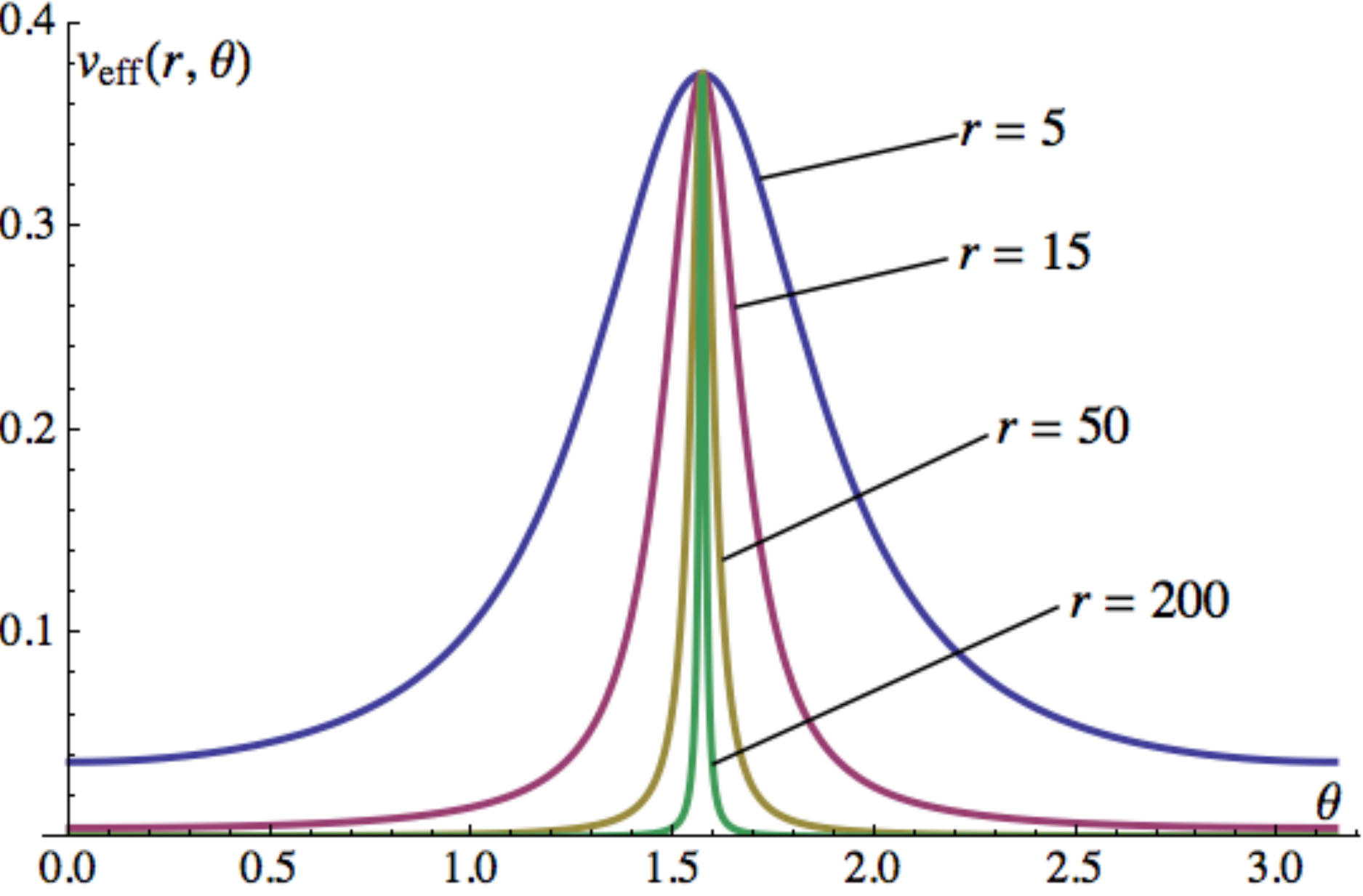}
   \caption{The effective potential $v_{\rm eff}(\br)$ for $\sqrt{n(\br)}$ in the case of $N=3$ non-interacting electrons in the external harmonic potential $v_{\rm ext}(\br)=\frac{1}{2}\omega^2r^2$ (with $\omega=\frac{1}{4}$) in the configuration in which the HOMO is a $p_z$ orbital. Top panel: $v_{\rm eff}(\br)$ as a function of $r=|\br|$ for different values of $\theta=\arccos(\frac{z}{r})$. Bottom panel: $v_{\rm eff}(\br)$ as a function of $\theta$ for different values of $r$. }
\label{fig:veff3eleharm}
\end{figure}

We first consider again $N=3$ non-interacting electrons and we put them in the harmonic potential $v_{\rm ext}(\br)=\frac{1}{2}\omega^2r^2$. With the lowest totally symmetric orbital ($s$ type) doubly occupied, and one electron available for the degenerate $p$ orbitals (configuration $s^2p^1$), we select among the three degenerate ground states the one with $m=0$ so that the HOMO is again a $p_z$ orbital.
The corresponding $v_{\rm eff}(\br)$ is calculated as in the previous section, see \eqref{eq:veffbyinversion}, and is reported in Fig.~\ref{fig:veff3eleharm}. We see that, on the HNP, $v_{\rm eff}(r_p\to\infty)$ again does not go to zero, but this time it tends to a constant. It is easy to verify analytically that if, asymptotically, $\psi_H\sim r^{q_0} f(\cos\theta) e^{-\frac{\omega}{2}r^2}$ with $f(0)=0$, and $\psi_{H-1}\sim  r^{q_1}e^{-\frac{\omega}{2}r^2}$, we have
\begin{eqnarray}
v^{kin}(r_p\to\infty) & \sim & f'(0)^2 \, r^{2(q_0-q_1-1)},  \nonumber \\
   q_0-q_1 & = & \frac{E_1^{N-1}-E_0^{N-1}}{\omega}.
\label{eq:vkinharm}
\end{eqnarray}
For non-interacting electrons we have always $E_1^{N-1}-E_0^{N-1}=\omega$ so that the potential goes to a constant in the plane,
\begin{align}
v_{\rm eff}(r_p\to\infty)&=v_{kin}(\infty) + v^{N-1}(\infty) \notag. \\
                                     &= v^{kin}(\infty)+E_1^{N-1}-E_0^{N-1}.
\end{align}
For $N\ge 3$ \textit{interacting} electrons, depending on how correlated is the system, we could have $E_1^{N-1}-E_0^{N-1}>\omega$, and thus a polynomially diverging behavior of $v_{\rm eff}$ on the HNP.
Comparison of Eqs.~\eqref{eq:vkinharm} and \eqref{eq:vkinasym} shows that in the presence of a HOMO nodal plane that extends to infinity the asymptotic behavior of $v_{\rm eff}(\br)$ on the plane can depend dramatically on the kind of binding external potential.  

\subsection{$N=2$ spin-polarized interacting electrons}\label{sec_harminteract}
We consider now $N=2$ spin-polarized interacting electrons (with standard Coulomb $1/r_{12}$ interaction). As well known, the corresponding hamiltonian is separable into center-of-mass ${\bf R}=\frac{1}{2}(\br_1+\br_2)$ and relative $\br_{12}=\br_2-\br_1$ coordinates, so that its exact wavefunction reads $\Psi_0^N(\br_1,\br_2)=\xi({\bf R})\phi(\br_{12})$.  With spin-polarized electrons, the spatial wavefunction must satisfy $\Psi_0^N(\br_1,\br_2)=-\Psi_0^N(\br_2,\br_1)$, which implies that the ground state corresponds to the $\ell_{12}=1$ spherical harmonic for the relative vector $\br_{12}$. We have then 3 degenerate ground-state wavefunctions, and we choose one of them by fixing $m_{12}=0$: this way, we obtain an interacting density with a symmetry plane like the one encountered in molecules.

For $N=2$ there is an infinite set of special values of $\omega$ for which $q$ in Eq.~\eqref{eq:asymptHarm} is integer: they correspond to analytical solutions of the interacting hamiltonian \cite{Taut1993}. For $\ell_{12}=1$, $\omega=\frac{1}{4}$ is one of those. The interacting wave function for the case $m_{12}=0$ then is equal to
\begin{equation}
	\label{eq:Psi0}
	\Psi_0^{N=2}(\br_1,\br_2)=C\,e^{-\frac{1}{8}(r_1^2+r_2^2)}(z_2-z_1)\left(1+\frac{|\br_2-\br_1|}{4}\right),
\end{equation}
with $C$ a normalization constant. The associated density is given by
\begin{eqnarray}
	n(\br) & = & C_n e^{-\frac{r^2}{4}}\Biggl\{\frac{\pi^{3/2}}{2}\biggl[2\left( 26+ r^2-z^2\right)+z^2\big(32 + r^2\big)\biggr]\nonumber \\
& + & \frac{4 \pi}{r^5}\biggl[e^{-\frac{r^2}{4}}\big(-24\, r\, z^2+8 r^3\big(1+z^2\big)+  2 r^5\big(2+z^2\big)\big) \nonumber \\ 
&+ &\sqrt{\pi}\,{\rm erf}\left(\frac{r}{2}\right)\biggl(24 z^2+r^6(2+z^2)-4 r^2\big(2+3 z^2\big) \nonumber \\
& + & r^4\big(8+6 z^2\big)\biggr)  \biggr]\Biggr\},
\label{eq:densN2int}
\end{eqnarray}
with the normalization constant
\begin{equation}
	C_n=\frac{3}{16\sqrt{2}\,\pi^{\frac{5}{2}}\,\left(64+27\,\sqrt{2\pi}\right)}.
\end{equation}
\begin{figure}
   \includegraphics[width=8.cm]{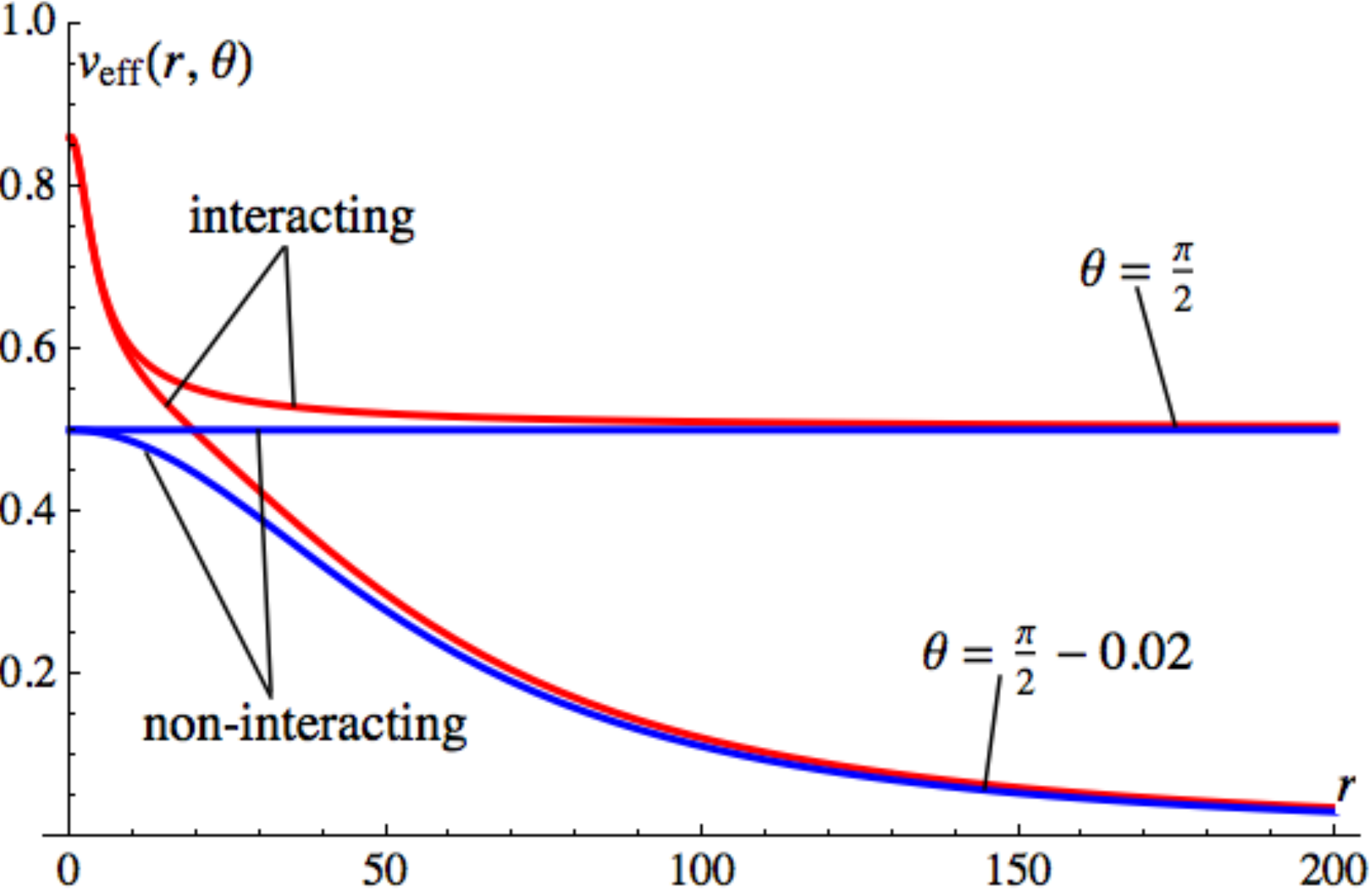}
   \caption{The effective potential $v_{\rm eff}(\br)$ in the case of $N=2$ spin-polarized electrons in the external potential $v_{\rm ext}(\br)=\frac{1}{2}\omega^2r^2$ (with $\omega=\frac{1}{4}$) as a function of $r=|\br|$ for two different values of $\theta=\arccos(\frac{z}{r})$.
The result for both interacting and non-interacting electrons is reported.}
\label{fig:veffHarmonic}
\end{figure}

Inserting Eq.~\eqref{eq:densN2int} into Eq.~\eqref{eq:veffbyinversion} we find that the corresponding $v_{\rm eff}(\br)$ has the same asymptotic behavior as observed for non-interacting electrons in the harmonic potential in Fig.~\ref{fig:veff3eleharm}. This is shown in Fig.~\ref{fig:veffHarmonic}, where we compare our $v_{\rm eff}(\br)$ for interacting electrons with the one for two non-interacting spin-polarized electrons in the same external potential. We clearly see that, in this case, the asymptotic behavior close to the HNP is exactly the same. This is in agreement with our findings of Sec.~\ref{sec:veffasymptotics} and \ref{sec_harmnoninter}: the behavior close to the HNP is entirely determined by the differences between the ground and the first excited states of the $N-1$ state. For $N=2$, the $N-1$ states are the same for both interacting and non-interacting electrons. 

In this case we can also compute analytically the first two Dyson orbitals that can be obtained from
\begin{equation}
	d_i(\br)=\sqrt{2}\int \Psi_i^{N=1}(\br')\Psi_0^{N=2}(\br,\br')d\br',
\end{equation}
with $\Psi_0^{N=2}(\br,\br')$ given by Eq.~\eqref{eq:Psi0}, and are reported in Appendix~\ref{app_dyson}. We see that, as considered in our discussion, we have
\begin{equation}
	d_0(|\br|\to\infty)\sim z\,r\,e^{-\frac{r^2}{8}},
\end{equation}
and
\begin{equation}
	d_1(|\br|\to\infty)\sim \left(r+\frac{z^2}{r}\right)\,e^{-\frac{r^2}{8}}.
\end{equation}
Notice that in this interacting case the second Dyson orbital has a weight slightly larger than the first one,
\begin{eqnarray}
	\int |d_0(\br)|^2\,d\br & = & 0.979516 \\
		\int |d_1(\br)|^2\,d\br & = &  0.98742.
\end{eqnarray}

\section{Summary and Conclusions}\label{Sec_Conclusions}
It is known that in Coulombic potentials (in atoms and molecules) there is an intimate relation between the asymptotic decay of the electron density and the first ionization potential.The exact density of an interacting electron system in such a potential does not always have a uniform asymptotic decay, but can carry different ionization-energy information in directions where the KS HOMO and the first Dyson orbital have a nodal plane.  We have earlier investigated the implications for the Kohn-Sham potential of DFT \cite{GoriGiorgiGalBaerends2016}. Here we investigated  the effective potential for $\sqrt{n}$ (and thus the functional derivative of the von Weizs\"acker kinetic energy functional) in the special case of a density which is represented by KS orbitals with a nodal plane in the KS HOMO and uniform asymptotic decay of HOMO$-1$ (a Case 1 density). Irrespective of the fact that the corresponding true density may not exhibit these precise features (see the Case 2 discussion), this type of density (the minimal model of Ref.~\cite{Aschebrock2017}) will occur often in regular KS calculations.   The effective potential for the density, $v_{\rm eff}(\br)$, will in that case deviate from the usual uniform asymptotic decay like $-1/r$, and  will diverge asymptotically.

We have also investigated the issue in the harmonic external potential and reported an interacting case that can be solved analytically in which the density on the nodal plane decays differently, supporting with the exact first two Dyson orbitals many of the assumptions used in our previous derivations. In future works we plan to study cases in which the ground-state wavefunction is complex.

\section*{Acknowledgments}
It is a pleasure to dedicate this paper to Hardy Gross, who has done excellent and inspiring work on the conditional amplitude formalism used here.\\
Financial support was provided by the European Research Council under H2020/ERC Consolidator Grant corr-DFT [Grant Number 648932]. 

\appendix
\section{The first two Dyson orbitals for the interacting spin-polarized harmonium atom}
\label{app_dyson}
By direct integrating the interacting wavefunction with the $N-1$ ground state and first excited state we obtain
\begin{widetext}
\begin{equation}
	d_0(\br)=C_0 z e^{-\frac{r^2}{8}}\left(8 \pi^{3/2}+\frac{2 \pi}{r^3}\left(2 e^{-\frac{r^2}{4}}(2 r+r^3)+\sqrt{\pi}(-4+4r^2+r^4){\rm erf}\left(\frac{r}{2}\right)\right)\right)
\end{equation}
\begin{multline}
	d_1(\br)=C_1 e^{-\frac{r^2}{8}}\biggl(16 \pi^{3/2}+\frac{4 \pi}{r^3}\left(2 e^{-\frac{r^2}{4}}(2 r+r^3)+\sqrt{\pi}(-4+4r^2+r^4){\rm erf}\left(\frac{r}{2}\right)\right)+ \\
	\frac{z^2}{r^5}4\pi\left(2r e^{-\frac{r^2}{4}}(r^2-6 )+
	\sqrt{\pi}(r^4-4 r^2+12){\rm erf}\left(\frac{r}{2}\right)\right)\biggr),
\end{multline}
\end{widetext}
with
\begin{eqnarray}
	C_0 & = & \frac{1}{8\pi^{3/4}}\sqrt{\frac{3}{2\sqrt{2}\pi^{5/2}\left(64+27\sqrt{\pi}\right)}} \\
	C_1 & = & \frac{\sqrt{2}}{16\pi^{3/4}}\sqrt{\frac{3}{2\sqrt{2}\pi^{5/2}\left(64+27\sqrt{\pi}\right)}}.
\end{eqnarray}


\begin{thebibliography}{35}
\expandafter\ifx\csname natexlab\endcsname\relax\def\natexlab#1{#1}\fi
\expandafter\ifx\csname bibnamefont\endcsname\relax
  \def\bibnamefont#1{#1}\fi
\expandafter\ifx\csname bibfnamefont\endcsname\relax
  \def\bibfnamefont#1{#1}\fi
\expandafter\ifx\csname citenamefont\endcsname\relax
  \def\citenamefont#1{#1}\fi
\expandafter\ifx\csname url\endcsname\relax
  \def\url#1{\texttt{#1}}\fi
\expandafter\ifx\csname urlprefix\endcsname\relax\def\urlprefix{URL }\fi
\providecommand{\bibinfo}[2]{#2}
\providecommand{\eprint}[2][]{\url{#2}}

\bibitem[{\citenamefont{Hunter}(1975{\natexlab{a}})}]{Hunter1975Symp9}
\bibinfo{author}{\bibfnamefont{G.}~\bibnamefont{Hunter}},
  \bibinfo{journal}{Intern. J. Quantum Chem. Symp.}
  \textbf{\bibinfo{volume}{9}}, \bibinfo{pages}{311}
  (\bibinfo{year}{1975}{\natexlab{a}}).

\bibitem[{\citenamefont{Levy et~al.}(1984)\citenamefont{Levy, Perdew, and
  Sahni}}]{LevyPerdewSahni1984}
\bibinfo{author}{\bibfnamefont{M.}~\bibnamefont{Levy}},
  \bibinfo{author}{\bibfnamefont{J.~P.} \bibnamefont{Perdew}},
  \bibnamefont{and} \bibinfo{author}{\bibfnamefont{V.}~\bibnamefont{Sahni}},
  \bibinfo{journal}{Phys. Rev. A} \textbf{\bibinfo{volume}{30}},
  \bibinfo{pages}{2745} (\bibinfo{year}{1984}).

\bibitem[{\citenamefont{Abedi et~al.}(2010)\citenamefont{Abedi, Maitra, and
  Gross}}]{AbeMaiGro-PRL-10}
\bibinfo{author}{\bibfnamefont{A.}~\bibnamefont{Abedi}},
  \bibinfo{author}{\bibfnamefont{N.~T.} \bibnamefont{Maitra}},
  \bibnamefont{and} \bibinfo{author}{\bibfnamefont{E.~K.~U.}
  \bibnamefont{Gross}}, \bibinfo{journal}{Phys. Rev. Lett.}
  \textbf{\bibinfo{volume}{105}}, \bibinfo{pages}{123002}
  (\bibinfo{year}{2010}).

\bibitem[{\citenamefont{Schild and Gross}(2017)}]{SchGro-PRL-17}
\bibinfo{author}{\bibfnamefont{A.}~\bibnamefont{Schild}} \bibnamefont{and}
  \bibinfo{author}{\bibfnamefont{E.~K.~U.} \bibnamefont{Gross}},
  \bibinfo{journal}{Phys. Rev. Lett.} \textbf{\bibinfo{volume}{118}},
  \bibinfo{pages}{163202} (\bibinfo{year}{2017}).

\bibitem[{\citenamefont{Della~Sala and
  G\"orling}(2002{\natexlab{a}})}]{DellaSalaGoerling2002}
\bibinfo{author}{\bibfnamefont{F.}~\bibnamefont{Della~Sala}} \bibnamefont{and}
  \bibinfo{author}{\bibfnamefont{A.}~\bibnamefont{G\"orling}},
  \bibinfo{journal}{Phys. Rev. Lett.} \textbf{\bibinfo{volume}{89}},
  \bibinfo{pages}{033003} (\bibinfo{year}{2002}{\natexlab{a}}).

\bibitem[{\citenamefont{Della~Sala and
  G\"orling}(2002{\natexlab{b}})}]{DellaSalaGoerling2002b}
\bibinfo{author}{\bibfnamefont{F.}~\bibnamefont{Della~Sala}} \bibnamefont{and}
  \bibinfo{author}{\bibfnamefont{A.}~\bibnamefont{G\"orling}},
  \bibinfo{journal}{J. Chem. Phys.} \textbf{\bibinfo{volume}{116}},
  \bibinfo{pages}{5374} (\bibinfo{year}{2002}{\natexlab{b}}).

\bibitem[{\citenamefont{K\"ummel and
  Perdew}(2003{\natexlab{a}})}]{KuemmelPerdew2003}
\bibinfo{author}{\bibfnamefont{S.}~\bibnamefont{K\"ummel}} \bibnamefont{and}
  \bibinfo{author}{\bibfnamefont{J.~P.} \bibnamefont{Perdew}},
  \bibinfo{journal}{Phys. Rev. Lett.} \textbf{\bibinfo{volume}{90}},
  \bibinfo{pages}{043004} (\bibinfo{year}{2003}{\natexlab{a}}).

\bibitem[{\citenamefont{K\"ummel and
  Perdew}(2003{\natexlab{b}})}]{KuemmelPerdew2003b}
\bibinfo{author}{\bibfnamefont{S.}~\bibnamefont{K\"ummel}} \bibnamefont{and}
  \bibinfo{author}{\bibfnamefont{J.~P.} \bibnamefont{Perdew}},
  \bibinfo{journal}{Phys. Rev. B} \textbf{\bibinfo{volume}{68}},
  \bibinfo{pages}{035103} (\bibinfo{year}{2003}{\natexlab{b}}).

\bibitem[{\citenamefont{Wu et~al.}(2003)\citenamefont{Wu, Ayers, and
  Yang}}]{WuAyersYang2003}
\bibinfo{author}{\bibfnamefont{Q.}~\bibnamefont{Wu}},
  \bibinfo{author}{\bibfnamefont{P.~W.} \bibnamefont{Ayers}}, \bibnamefont{and}
  \bibinfo{author}{\bibfnamefont{W.}~\bibnamefont{Yang}}, \bibinfo{journal}{J.
  Chem. Phys.} \textbf{\bibinfo{volume}{119}}, \bibinfo{pages}{2978}
  (\bibinfo{year}{2003}).

\bibitem[{\citenamefont{Gori-Giorgi et~al.}(2016)\citenamefont{Gori-Giorgi,
  G\'al, and Baerends}}]{GoriGiorgiGalBaerends2016}
\bibinfo{author}{\bibfnamefont{P.}~\bibnamefont{Gori-Giorgi}},
  \bibinfo{author}{\bibfnamefont{T.}~\bibnamefont{G\'al}}, \bibnamefont{and}
  \bibinfo{author}{\bibfnamefont{E.~J.} \bibnamefont{Baerends}},
  \bibinfo{journal}{Mol. Phys.} \textbf{\bibinfo{volume}{114}},
  \bibinfo{pages}{1086} (\bibinfo{year}{2016}).

\bibitem[{\citenamefont{Aschebrock et~al.}(2017)\citenamefont{Aschebrock,
  Armiento, and K\"ummel}}]{Aschebrock2017}
\bibinfo{author}{\bibfnamefont{T.}~\bibnamefont{Aschebrock}},
  \bibinfo{author}{\bibfnamefont{R.}~\bibnamefont{Armiento}}, \bibnamefont{and}
  \bibinfo{author}{\bibfnamefont{S.}~\bibnamefont{K\"ummel}},
  \bibinfo{journal}{Phys. Rev. B} \textbf{\bibinfo{volume}{95}},
  \bibinfo{pages}{245118} (\bibinfo{year}{2017}).

\bibitem[{\citenamefont{{von Weizs\"acker}}(1935)}]{Wei-ZP-35}
\bibinfo{author}{\bibfnamefont{C.~F.} \bibnamefont{{von Weizs\"acker}}},
  \bibinfo{journal}{Z. Phys.} \textbf{\bibinfo{volume}{{96}}},
  \bibinfo{pages}{431} (\bibinfo{year}{1935}).

\bibitem[{\citenamefont{March}(1986)}]{Mar-PLA-86}
\bibinfo{author}{\bibfnamefont{N.~H.} \bibnamefont{March}},
  \bibinfo{journal}{Phys. Lett.} \textbf{\bibinfo{volume}{{113A}}},
  \bibinfo{pages}{476} (\bibinfo{year}{1986}).

\bibitem[{\citenamefont{Levy and Ou-Yang}(1988)}]{LevOu--PRA-88}
\bibinfo{author}{\bibfnamefont{M.}~\bibnamefont{Levy}} \bibnamefont{and}
  \bibinfo{author}{\bibfnamefont{H.}~\bibnamefont{Ou-Yang}},
  \bibinfo{journal}{Phys. Rev. A} \textbf{\bibinfo{volume}{{38}}},
  \bibinfo{pages}{625} (\bibinfo{year}{1988}).

\bibitem[{\citenamefont{Holas and March}(1991)}]{HolMar-PRA-91}
\bibinfo{author}{\bibfnamefont{A.}~\bibnamefont{Holas}} \bibnamefont{and}
  \bibinfo{author}{\bibfnamefont{N.~H.} \bibnamefont{March}},
  \bibinfo{journal}{Phys. Rev. A} \textbf{\bibinfo{volume}{{44}}},
  \bibinfo{pages}{5521} (\bibinfo{year}{1991}).

\bibitem[{\citenamefont{Ligeneres and Carter}(2005)}]{LigCar-HMM-05}
\bibinfo{author}{\bibfnamefont{V.~L.} \bibnamefont{Ligeneres}}
  \bibnamefont{and} \bibinfo{author}{\bibfnamefont{E.~A.}
  \bibnamefont{Carter}}, in \emph{\bibinfo{booktitle}{Handbook of Materials
  Modeling}}, edited by \bibinfo{editor}{\bibfnamefont{S.}~\bibnamefont{Yip}}
  (\bibinfo{publisher}{Springer}, \bibinfo{address}{The Netherlands},
  \bibinfo{year}{2005}), pp. \bibinfo{pages}{137--148}.

\bibitem[{\citenamefont{Becke}(1993)}]{Becke1993}
\bibinfo{author}{\bibfnamefont{A.~D.} \bibnamefont{Becke}},
  \bibinfo{journal}{J. Chem. Phys.} \textbf{\bibinfo{volume}{98}},
  \bibinfo{pages}{5648} (\bibinfo{year}{1993}).

\bibitem[{\citenamefont{Becke}(1996)}]{Bec-JCP-96}
\bibinfo{author}{\bibfnamefont{A.~D.} \bibnamefont{Becke}},
  \bibinfo{journal}{J. Chem. Phys.} \textbf{\bibinfo{volume}{104}},
  \bibinfo{pages}{1040} (\bibinfo{year}{1996}).

\bibitem[{\citenamefont{Tao et~al.}(2003)\citenamefont{Tao, Perdew, Staroverov,
  and Scuseria}}]{TaoPerStaScu-PRL-03}
\bibinfo{author}{\bibfnamefont{J.}~\bibnamefont{Tao}},
  \bibinfo{author}{\bibfnamefont{J.~P.} \bibnamefont{Perdew}},
  \bibinfo{author}{\bibfnamefont{V.~N.} \bibnamefont{Staroverov}},
  \bibnamefont{and} \bibinfo{author}{\bibfnamefont{G.~E.}
  \bibnamefont{Scuseria}}, \bibinfo{journal}{Phys. Rev. Lett.}
  \textbf{\bibinfo{volume}{{91}}}, \bibinfo{pages}{146401}
  (\bibinfo{year}{2003}).

\bibitem[{\citenamefont{Perdew et~al.}(2005)\citenamefont{Perdew, Ruzsinszky,
  Tao, Staroverov, Scuseria, and Csonka}}]{PerRuzTaoStaScuCso-JCP-05}
\bibinfo{author}{\bibfnamefont{J.~P.} \bibnamefont{Perdew}},
  \bibinfo{author}{\bibfnamefont{A.}~\bibnamefont{Ruzsinszky}},
  \bibinfo{author}{\bibfnamefont{J.}~\bibnamefont{Tao}},
  \bibinfo{author}{\bibfnamefont{V.~N.} \bibnamefont{Staroverov}},
  \bibinfo{author}{\bibfnamefont{G.~E.} \bibnamefont{Scuseria}},
  \bibnamefont{and} \bibinfo{author}{\bibfnamefont{G.~I.}
  \bibnamefont{Csonka}}, \bibinfo{journal}{J. Chem. Phys.}
  \textbf{\bibinfo{volume}{123}}, \bibinfo{pages}{062201}
  (\bibinfo{year}{2005}).

\bibitem[{\citenamefont{Zhao et~al.}(2006)\citenamefont{Zhao, Schultz, and
  Truhlar}}]{ZhaSchTru-JCTC-06}
\bibinfo{author}{\bibfnamefont{Y.}~\bibnamefont{Zhao}},
  \bibinfo{author}{\bibfnamefont{N.~E.} \bibnamefont{Schultz}},
  \bibnamefont{and} \bibinfo{author}{\bibfnamefont{D.~G.}
  \bibnamefont{Truhlar}}, \bibinfo{journal}{J. Chem. Theory Comput.}
  \textbf{\bibinfo{volume}{2}}, \bibinfo{pages}{364} (\bibinfo{year}{2006}).

\bibitem[{\citenamefont{Perdew et~al.}(2009)\citenamefont{Perdew, Ruzsinszky,
  Csonka, Constantin, and Sun}}]{PerRuzCsoConSun-PRL-09}
\bibinfo{author}{\bibfnamefont{J.~P.} \bibnamefont{Perdew}},
  \bibinfo{author}{\bibfnamefont{A.}~\bibnamefont{Ruzsinszky}},
  \bibinfo{author}{\bibfnamefont{G.~I.} \bibnamefont{Csonka}},
  \bibinfo{author}{\bibfnamefont{L.~A.} \bibnamefont{Constantin}},
  \bibnamefont{and} \bibinfo{author}{\bibfnamefont{J.}~\bibnamefont{Sun}},
  \bibinfo{journal}{Phys. Rev. Lett.} \textbf{\bibinfo{volume}{{103}}},
  \bibinfo{pages}{026403} (\bibinfo{year}{2009}).

\bibitem[{\citenamefont{Taut}(1993)}]{Taut1993}
\bibinfo{author}{\bibfnamefont{M.}~\bibnamefont{Taut}}, \bibinfo{journal}{Phys.
  Rev. A} \textbf{\bibinfo{volume}{48}}, \bibinfo{pages}{3561}
  (\bibinfo{year}{1993}).

\bibitem[{\citenamefont{Katriel and Davidson}(1980)}]{KatrielDavidson1980}
\bibinfo{author}{\bibfnamefont{J.}~\bibnamefont{Katriel}} \bibnamefont{and}
  \bibinfo{author}{\bibfnamefont{E.~R.} \bibnamefont{Davidson}},
  \bibinfo{journal}{Proc. Natl. Acad. Sci. USA} \textbf{\bibinfo{volume}{77}},
  \bibinfo{pages}{4403} (\bibinfo{year}{1980}).

\bibitem[{\citenamefont{Hunter}(1975{\natexlab{b}})}]{Hunter1975}
\bibinfo{author}{\bibfnamefont{G.}~\bibnamefont{Hunter}},
  \bibinfo{journal}{Intern. J. Quantum Chem.} \textbf{\bibinfo{volume}{9}},
  \bibinfo{pages}{237} (\bibinfo{year}{1975}{\natexlab{b}}).

\bibitem[{\citenamefont{Handy et~al.}(1969)\citenamefont{Handy, Marron, and
  Silverstone}}]{HandyMarronSilverstone1969}
\bibinfo{author}{\bibfnamefont{N.~C.} \bibnamefont{Handy}},
  \bibinfo{author}{\bibfnamefont{M.~T.} \bibnamefont{Marron}},
  \bibnamefont{and} \bibinfo{author}{\bibfnamefont{H.~J.}
  \bibnamefont{Silverstone}}, \bibinfo{journal}{Phys. Rev.}
  \textbf{\bibinfo{volume}{180}}, \bibinfo{pages}{45} (\bibinfo{year}{1969}).

\bibitem[{\citenamefont{Holas}(2008)}]{Holas2008}
\bibinfo{author}{\bibfnamefont{A.}~\bibnamefont{Holas}},
  \bibinfo{journal}{Phys. Rev. A.} \textbf{\bibinfo{volume}{77}},
  \bibinfo{pages}{026501} (\bibinfo{year}{2008}).

\bibitem[{\citenamefont{Armiento and K\"ummel}(2013)}]{ArmientoKuemmel2013}
\bibinfo{author}{\bibfnamefont{R.}~\bibnamefont{Armiento}} \bibnamefont{and}
  \bibinfo{author}{\bibfnamefont{S.}~\bibnamefont{K\"ummel}},
  \bibinfo{journal}{Phys. Rev. Lett.} \textbf{\bibinfo{volume}{111}},
  \bibinfo{pages}{036402} (\bibinfo{year}{2013}).

\bibitem[{\citenamefont{Becke}(1988)}]{Becke1988}
\bibinfo{author}{\bibfnamefont{A.~D.} \bibnamefont{Becke}},
  \bibinfo{journal}{Phys. Rev. A} \textbf{\bibinfo{volume}{38}},
  \bibinfo{pages}{3098} (\bibinfo{year}{1988}).

\bibitem[{\citenamefont{Becke and Johnson}(2006)}]{BeckeJohnson2006}
\bibinfo{author}{\bibfnamefont{A.~D.} \bibnamefont{Becke}} \bibnamefont{and}
  \bibinfo{author}{\bibfnamefont{E.~R.} \bibnamefont{Johnson}},
  \bibinfo{journal}{J. Chem. Phys.} \textbf{\bibinfo{volume}{124}},
  \bibinfo{pages}{221101} (\bibinfo{year}{2006}).

\bibitem[{\citenamefont{van Leeuwen and Baerends}(1994)}]{LeeuwenBaerends1994}
\bibinfo{author}{\bibfnamefont{R.}~\bibnamefont{van Leeuwen}} \bibnamefont{and}
  \bibinfo{author}{\bibfnamefont{E.~J.} \bibnamefont{Baerends}},
  \bibinfo{journal}{Phys. Rev. A} \textbf{\bibinfo{volume}{49}},
  \bibinfo{pages}{2421} (\bibinfo{year}{1994}).

\bibitem[{\citenamefont{Buijse et~al.}(1989)\citenamefont{Buijse, Baerends, and
  Snijders}}]{BuijseBaerendsSnijders1989}
\bibinfo{author}{\bibfnamefont{M.~A.} \bibnamefont{Buijse}},
  \bibinfo{author}{\bibfnamefont{E.~J.} \bibnamefont{Baerends}},
  \bibnamefont{and} \bibinfo{author}{\bibfnamefont{J.~G.}
  \bibnamefont{Snijders}}, \bibinfo{journal}{Phys. Rev. A}
  \textbf{\bibinfo{volume}{40}}, \bibinfo{pages}{4190} (\bibinfo{year}{1989}).

\bibitem[{\citenamefont{Gritsenko et~al.}(2003)\citenamefont{Gritsenko,
  Bra\"ida, and Baerends}}]{GritsenkoBraidaBaerends2003}
\bibinfo{author}{\bibfnamefont{O.~V.} \bibnamefont{Gritsenko}},
  \bibinfo{author}{\bibfnamefont{B.}~\bibnamefont{Bra\"ida}}, \bibnamefont{and}
  \bibinfo{author}{\bibfnamefont{E.~J.} \bibnamefont{Baerends}},
  \bibinfo{journal}{J. Chem. Phys.} \textbf{\bibinfo{volume}{119}},
  \bibinfo{pages}{1937} (\bibinfo{year}{2003}).

\bibitem[{\citenamefont{Baerends and
  Gritsenko}(1997)}]{BaerendsGritsenkoJPCA1997}
\bibinfo{author}{\bibfnamefont{E.~J.} \bibnamefont{Baerends}} \bibnamefont{and}
  \bibinfo{author}{\bibfnamefont{O.~V.} \bibnamefont{Gritsenko}},
  \bibinfo{journal}{J. Phys. Chem. A} \textbf{\bibinfo{volume}{101}},
  \bibinfo{pages}{5383} (\bibinfo{year}{1997}).

\bibitem[{\citenamefont{Chong et~al.}(2002)\citenamefont{Chong, Gritsenko, and
  Baerends}}]{ChongGritsenkoBaerends2002}
\bibinfo{author}{\bibfnamefont{D.~P.} \bibnamefont{Chong}},
  \bibinfo{author}{\bibfnamefont{O.~V.} \bibnamefont{Gritsenko}},
  \bibnamefont{and} \bibinfo{author}{\bibfnamefont{E.~J.}
  \bibnamefont{Baerends}}, \bibinfo{journal}{J. Chem. Phys.}
  \textbf{\bibinfo{volume}{116}}, \bibinfo{pages}{1760} (\bibinfo{year}{2002}).

\end{thebibliography}

\end{document}